\documentclass[prd,preprint,superscriptaddress,showpacs,byrevtex]{revtex4}
\usepackage{bm}
\def\fsl#1{\setbox0=\hbox{$#1$}           
   \dimen0=\wd0                                 
   \setbox1=\hbox{/} \dimen1=\wd1               
   \ifdim\dimen0>\dimen1                        
      \rlap{\hbox to \dimen0{\hfil/\hfil}}      
      #1                                        
   \else                                        
      \rlap{\hbox to \dimen1{\hfil$#1$\hfil}}   
      /                                         
   \fi}                                         %
\newcommand{\be}{\begin{equation}}
\newcommand{\ee}{\end{equation}}
\newcommand{\bea}{\begin{eqnarray}}
\newcommand{\eea}{\end{eqnarray}}
\newcommand{\beq}{\begin{equation}}
\newcommand{\eeq}{\end{equation}}
\newcommand{\beqs}{\begin{eqnarray}}
\newcommand{\eeqs}{\end{eqnarray}}

\begin{document}

\title{ Proof of Factorization of $\chi_{cJ}$ Production in Non-Equilibrium QCD at RHIC and LHC in Color Singlet Mechanism }
\author{Gouranga C Nayak }\thanks{G. C. Nayak was affiliated with C. N. Yang Institute for Theoretical Physics in 2004-2007.}
\affiliation{ C. N. Yang Institute for Theoretical Physics, Stony Brook University, Stony Brook NY, 11794-3840 USA}
%
\begin{abstract}
Recently we have proved the factorization of NRQCD S-wave heavy quarkonium production at all orders in coupling constant. In this paper we extend this to prove the factorization of infrared divergences in $\chi_{cJ}$ production from color singlet $c{\bar c}$ pair in non-equilibrium QCD at RHIC and LHC at all orders in coupling constant. This can be relevant to study the quark-gluon plasma at RHIC and LHC.
\end{abstract}
\maketitle
\pagestyle{plain}
\pagenumbering{arabic}
\section{Introduction}
The factorization of infrared divergences in nonrelativistic QCD (NRQCD) color octet S-wave heavy quarkonium production at high energy colliders at all orders in coupling constant is recently proved in \cite{nnr}. In this paper we extend this formalism to non-equilibrium QCD by using the closed-time path integral formulation to prove the factorization of infrared divergences in $\chi_{cJ}$ production from the color singlet $c{\bar c}$ pair in non-equilibrium QCD at all orders in coupling constant at RHIC and LHC. We also predict the correct definition of the non-perturbative matrix element of the $\chi_{cJ}$ production from color singlet $c{\bar c}$ pair in non-equilibrium QCD at RHIC and LHC. This can be relevant to study the quark-gluon plasma (QGP) at RHIC and LHC.

At very high temperature ($\ge$ 200 MeV) the normal hadronic matter becomes a new state of matter known as the QGP. About $10^{-12}$ seconds after the big bang our universe was filled with the QGP which makes it important to produce it in the laboratory at RHIC and LHC by colliding two heavy nuclei at very high energy \cite{nc}. Since the confinement in QCD prevents us to detect the QGP directly at RHIC and LHC, various indirect signatures (such as the heavy quarkonium production/suppression \cite{sz}) are proposed for its detection.

Since the center of mass energy ${\sqrt s}$ = 200 GeV (5.5 TeV) of Au-Au (Pb-Pb) collisions at RHIC (LHC) is
very high, the two nuclei at RHIC (LHC) travel almost at the speed of light creating the non-equilibrium quark-gluon plasma just after the heavy-ion collisions. Because of the very small hadronization time scale in QCD ($\sim 10^{-24}$ seconds) there may not be enough secondary partonic collisions to bring this non-equilibrium QGP to equilibrium. Hence the QGP at RHIC (LHC) may be in non-equilibrium where one can not define a temperature.

The hard (high $p_T$) parton production at RHIC and LHC can be calculated by using pQCD but the soft parton production calculation needs non-perturbative QCD which is not solved yet. This implies that there remains uncertainty in determining the soft partons production at RHIC and LHC. Note that the soft partons play an important role in determining the bulk properties of the QGP at RHIC and LHC.

It should be mentioned here that the study of hadronization from non-equilibrium QGP at RHIC and LHC is one of the most difficult and important problem because the
confinement problem in QCD is not solved yet due to the lack of our understanding of non-perturbative QCD. This implies that the first principle calculation of hadron production from non-equilibrium partons at RHIC and LHC is not known.

Because of these reasons one finds that in order to detect the QGP at RHIC and LHC by using the first principle calculation one needs to study the nonequilibrium-nonperturbative QCD by using the closed-time path integral formalism which is not easy \cite{s,k,g,c}. If one does not perform the exact first principle nonequilibrium-nonperturbative QCD calculation then the comparison of the theoretical calculation with the experimental data at RHIC and LHC becomes questionable. For example, some of the limitations of the present theoretical approaches are listed below.

The lattice QCD at finite temperature \cite{lt} is a common tool to study the properties of the QGP. However, for the reasons explained above, the actual QGP at RHIC and LHC may be in non-equilibrium where one can not define a temperature. Hence the lattice QCD at finite temperature has no application in non-equilibrium QGP at RHIC and LHC.

Similarly the hydrodynamics \cite{hd} is not applicable in non-equilibrium QGP at RHIC and LHC. Another limitation of the hydrodynamics \cite{hd} is that it does not answer the question how the partons become hadrons from first principle. As shown in \cite{njq} the parton to hadron fragmentation function in QCD in vacuum can not be used to study the hadrons production from partons from the quark-gluon plasma at RHIC and LHC. It is important to observe that even if the experimental data at RHIC and LHC is explained by using the hydrodynamics \cite{hd} it does not prove that the QGP is in equilibrium. In order to make sure that the QGP is in equilibrium at RHIC and LHC one has to prove that the same experimental data can not be explained by using the non-equilibrium QGP for which one has to study the nonequilibrium-nonperturbative QCD by using the closed-time path integral formalism.

As far as the actual physics at RHIC and LHC heavy-ion collisions is concerned the AdS/CFT based studies \cite{ad} and the supersymmetric Yang-Mills plasma based studies \cite{sy} have nothing to do it because of the lack of experimental verification of the string theory and the supersymmetry.

Regarding the initial condition for the QGP formation and the color glass condensate (CGC) \cite{cg}, as discussed above, the hard (high $p_T$) parton production at RHIC and LHC can be calculated by using the pQCD but the soft parton production can only be correctly calculated from the first principle by using the non-perturbative QCD which is yet to be solved.

The jet quenching study, see for example \cite{gv,gv1}, directly/indirectly uses the parton to hadron fragmentation function in QCD in vacuum. This is not possible because unlike the leading order perturbative gluon propagator in non-equilibrium QCD the non-perturbative fragmentation function in non-equilibrium QCD can not be decomposed into the vacuum part and the medium part \cite{njq}.

Hence from the above discussions one finds that, although a lot of experimental data is available at RHIC and LHC heavy-ion colliders, but there exists no exact first principle theoretical calculation to explain these experimental data. It is almost impossible to make an exact first principle theoretical calculation at RHIC and LHC without studying the nonequilibrium-noperturbative QCD by using the closed-time path integral formalism.

The first principle way to study non-equilibrium quantum field theory is the Schwinger-Keldysh closed-time path (CTP) formalism \cite{s,k}. Although the  non-equilibrium QED is usually studied by using the canonical quantization formalism, the closed-time path integral formalism is useful to study the nonequilibrium-nonperturbative QCD due to the self gluon interactions and the hadronization.

As mentioned earlier, the heavy quarkonium is one of the indirect signature for the detection of QGP \cite{sz}. Both $j/\psi$ and $\chi_{cJ}$ are measured by various collaborations at the RHIC and LHC heavy-ion collider experiments. In order to study heavy quarkonium production from the QGP at RHIC and LHC one needs to prove factorization of infrared divergences, otherwise one will predict infinite cross section for the heavy quarkonium production.

The infrared divergences issue in the case of P-wave heavy quarkonium production is more complicated than that of the $j/\psi$ production. This is because there are no uncanceled infrared divergences due to eikonal gluons exchange in the case of S-wave heavy quarkonium ($j/\psi$) production in the color singlet mechanism whereas there are uncanceled infrared divergences due to eikonal gluons exchange in case of P-wave heavy quarkonium ($\chi_{cJ}$ ) production in the color singlet mechanism \cite{b}.

Recently we have shown that these uncanceled infrared divergences can be factored into the correct definition of the color singlet P-wave heavy quarkonium non-perturbative matrix element by supplying the eikonal lines or the gauge links \cite{np}. In this paper we will extend this to the non-equilibrium QCD by using the closed-time path integral formalism. We will prove the factorization of infrared divergences in the $\chi_{cJ}$ production from the color singlet $c{\bar c}$ pair in non-equilibrium QCD at RHIC and LHC at all orders in coupling constant. We will predict the correct definition of the non-perturbative matrix element of the $\chi_{cJ}$ production from the color singlet $c{\bar c}$ pair in non-equilibrium QCD at RHIC and LHC. This can be relevant to detect the QGP at RHIC and LHC.

The paper is organized as follows. In section II a brief discussion on the generating functional in non-equilibrium QCD is presented. In section III we discuss the non-canceling infrared divergences in color singlet $\chi_{cJ}$ production. In section IV we show that the infrared divergences due to eikonal gluons exchange can be studied by using the SU(3) pure gauge. In section V we prove the factorization of infrared divergences in the $\chi_{cJ}$ production from color singlet $c{\bar c}$ pair in non-equilibrium QCD at RHIC and LHC at all orders in coupling constant. In section VI we predict the correct definition of the non-perturbative matrix element of the $\chi_{cJ}$ production from color singlet $c{\bar c}$ pair in non-equilibrium QCD at RHIC and LHC. We conclude in section VII.

\section{ closed-time path integral formalism and the generating functional in non-equilibrium QCD}

Since we will use the background field method of QCD in this paper we denote the gluon field by $Q^{\lambda d}(x)$ and the background field by $A^{\lambda d}(x)$ where $\lambda=0,1,2,3$ and $d=1,...,8$. The generating functional in non-equilibrium QCD (without the background field) in the closed-time path integral formalism is given by \cite{g,c}
\bea
&&Z[\rho,J_+,J_-,\eta_{1+},{\bar \eta}_{1+},\eta_{1-},{\bar \eta}_{1-},\eta_{2+},{\bar \eta}_{2+},\eta_{2-},{\bar \eta}_{2-},\eta_{3+},{\bar \eta}_{3+},\eta_{3-},{\bar \eta}_{3-},\eta_{I+},{\bar \eta}_{I+},\eta_{I-},{\bar \eta}_{I-}] \nonumber \\
&&
=\int [dQ_+] [dQ_-]\Pi_{k=1}^3[d{\bar \psi}_{k+}] [d{\bar \psi}_{k-}] [d \psi_{k+} ] [d\psi_{k-}]~[d{\bar \Psi}_{+}] [d{\bar \Psi}_{-}] [d \Psi_{+} ] [d\Psi_{-}]\nonumber \\
&&
\times {\rm det}(\frac{\delta \partial_\lambda Q_+^{\lambda d}}{\delta \omega_+^e})\times {\rm det}(\frac{\delta \partial_\lambda Q_-^{\lambda d}}{\delta \omega_-^e}) {\rm exp}[i\int d^4x {\bf \{}-\frac{1}{4}{F^d}_{\lambda \delta}^2[Q_+]+\frac{1}{4}{F^d}_{\lambda \delta}^2[Q_-] -\frac{1}{2 \alpha}(\partial_\lambda Q_+^{\lambda d })^2+\frac{1}{2 \alpha} (\partial_\lambda Q_-^{\lambda d })^2\nonumber \\
&&+\sum_{k=1}^3{\bar \psi}_{k+}  [i\gamma^\lambda \partial_\lambda -m_k +gT^d\gamma^\lambda Q^d_{\lambda +}]  \psi_{k+} -\sum_{k=1}^3{\bar \psi}_{k-}  [i\gamma^\lambda \partial_\lambda -m_k +gT^d\gamma^\lambda Q^d_{\lambda -}]  \psi_{k-}\nonumber \\
&&+{\bar \Psi}_{+}  [i\gamma^\lambda \partial_\lambda -M +gT^d\gamma^\lambda Q^d_{\lambda +}]  \Psi_{+}-{\bar \Psi}_{-}  [i\gamma^\lambda \partial_\lambda -M +gT^d\gamma^\lambda Q^d_{\lambda -}]  \Psi_{-} +J_+Q_+-J_-Q_-\nonumber \\
&&+\sum_{k=1}^3 [{\bar \psi}_{k+}\eta_{k+} - {\bar \psi}_{k-}\eta_{k-}+ {\bar \eta}_{k+}\psi_{k+} - {\bar \eta}_{k-}\psi_{k-}]+{\bar \Psi}_{+}\eta_{I+} -{\bar \Psi}_{-}\eta_{I-}+{\bar \eta}_{I+}\Psi_{+} -{\bar \eta}_{I-}\Psi_{-}{\bf \}}]\nonumber \\
&&  \times <Q_+,\psi_{1+},{\bar \psi}_{1+},\psi_{2+},{\bar \psi}_{2+},\psi_{3+},{\bar \psi}_{3+},\Psi_+,{\bar \Psi}_+,0|~\rho~|0,{\bar \psi}_{1-},\psi_{1-},{\bar \psi}_{2-},\psi_{2-},{\bar \psi}_{3-},\psi_{3-},\nonumber \\
&&{\bar \Psi}_-,\Psi_-,Q_->
\label{zfnq}
\eea
where $\delta=0,1,2,3$ and we have included the heavy quark. In eq. (\ref{zfnq}) the symbol $k=1,2,3=u, d, s$ stands for up, down and strange quark with mass $m_k$ and field $\psi_k$. The heavy quark field is $\Psi$ and the heavy quark mass is $M$. The initial density of states is denoted by $\rho$, the arbitrary gauge fixing parameter is $\alpha$, the $|0,{\bar \psi}_{1-},\psi_{1-},{\bar \psi}_{2-},\psi_{2-},{\bar \psi}_{3-},\psi_{3-},{\bar \Psi}_-,\Psi_-,Q_->$ corresponds to the state at the initial time and
\bea
&&{F^d}_{\lambda \delta}^2[Q_+]=[\partial_\lambda Q_{\delta +}^d-\partial_\delta Q_{\lambda +}^d+gf^{dba} Q_{\lambda +}^bQ_{\delta +}^a] \times [\partial^\lambda Q^{\delta d}_+-\partial^\delta Q^{\lambda d }_++gf^{dce} Q^{\lambda c}_+Q^{\delta e}_+] \nonumber \\
\label{qf2}
\eea
and similarly for the $-$ index where $+,-$ stand for the closed-time path indices. Note that we do not introduce ghost fields as we directly work with the ghost determinant ${\rm det}(\frac{\delta \partial_\lambda Q_+^{\lambda d}}{\delta \omega_+^e})$ in eq. (\ref{zfnq}).

The corresponding non-equilibrium QCD generating functional in the closed-time path integral formalism of the background field method of QCD is given by \cite{g,c,t,a,z}
\bea
&&Z[A,\rho,J_+,J_-,\eta_{1+},{\bar \eta}_{1+},\eta_{1-},{\bar \eta}_{1-},\eta_{2+},{\bar \eta}_{2+},\eta_{2-},{\bar \eta}_{2-},\eta_{3+},{\bar \eta}_{3+},\eta_{3-},{\bar \eta}_{3-},\eta_{I+},{\bar \eta}_{I+},\eta_{I-},{\bar \eta}_{I-}] \nonumber \\
&&
=\int [dQ_+] [dQ_-]\Pi_{k=1}^3[d{\bar \psi}_{k+}] [d{\bar \psi}_{k-}] [d \psi_{k+} ] [d\psi_{k-}]~[d{\bar \Psi}_{+}] [d{\bar \Psi}_{-}] [d \Psi_{+} ] [d\Psi_{-}]\nonumber \\
&& \times {\rm det}(\frac{\delta G^d( Q_+)}{\delta \omega_+^e})\times {\rm det}(\frac{\delta G^d(Q_-)}{\delta \omega_-^e}) \nonumber \\
&&\times {\rm exp}[i\int d^4x {\bf \{}-\frac{1}{4}{F^d}_{\lambda \delta}^2[Q_++A_+]+\frac{1}{4}{F^d}_{\lambda \delta}^2[Q_-+A_-] -\frac{1}{2 \alpha}(G^d(Q_+))^2+\frac{1}{2 \alpha} (G^d(Q_-))^2\nonumber \\
&&+\sum_{k=1}^3{\bar \psi}_{k+}  [i\gamma^\lambda \partial_\lambda -m_k +gT^d\gamma^\lambda (Q+A)^d_{\lambda +}]  \psi_{k+} -\sum_{k=1}^3{\bar \psi}_{k-}  [i\gamma^\lambda \partial_\lambda -m_k +gT^d\gamma^\lambda (Q+A)^d_{\lambda -}]  \psi_{k-}\nonumber \\
&&+{\bar \Psi}_{+}  [i\gamma^\lambda \partial_\lambda -M +gT^d\gamma^\lambda (Q+A)^d_{\lambda +}]  \Psi_{+}-{\bar \Psi}_{-}  [i\gamma^\lambda \partial_\lambda -M +gT^d\gamma^\lambda (Q+A)^d_{\lambda -}]  \Psi_{-}+\sum_{k=1}^3 [{\bar \psi}_{k+}\eta_{k+} \nonumber \\
&&- {\bar \psi}_{k-}\eta_{k-}+ {\bar \eta}_{k+}\psi_{k+} - {\bar \eta}_{k-}\psi_{k-}]+{\bar \Psi}_{+}\eta_{I+} -{\bar \Psi}_{-}\eta_{I-}+{\bar \eta}_{I+}\Psi_{+} -{\bar \eta}_{I-}\Psi_{-}+J_+Q_+-J_-Q_-{\bf \}}]\nonumber \\
&& \times <Q_++A_+,\psi_{1+},{\bar \psi}_{1+},\psi_{2+},{\bar \psi}_{2+},\psi_{3+},{\bar \psi}_{3+},\Psi_+,{\bar \Psi}_+,0|~\rho~|0,{\bar \psi}_{1-},\psi_{1-},{\bar \psi}_{2-},\psi_{2-},{\bar \psi}_{3-}\nonumber \\
&&,\psi_{3-},{\bar \Psi}_-,\Psi_-,Q_-+A_->
\label{zfb}
\eea
where the background gauge fixing
\bea
&& G^d(Q_+) =\partial_\lambda Q^{\lambda d}_+ + gf^{dba} A_{\lambda +}^b Q^{\lambda a}_+
\label{zgf}
\eea
depends on the background field $A^{\lambda d}(x)$. In eq. (\ref{zfb})
\bea
&& {F^d}_{\lambda \delta}^2[Q_++A_+]=[\partial_\lambda [A_{\delta +}^d+Q_{\delta +}^d]-\partial_\delta [A_{\lambda +}^d+Q_{\lambda +}^d]+gf^{dba} [A_{ \lambda +}^b+Q_{\lambda +}^b][A_{\delta +}^a+Q_{\delta +}^a]]\nonumber \\
&& \times [\partial^\lambda [A^{\delta d}_++Q^{\delta d}_+]-\partial^\delta [A^{\lambda d}_++Q^{\lambda d}_+]+gf^{dce} [A^{ \lambda c}_++Q^{\lambda c}_+][A^{\delta e}_++Q^{\delta e}_+]]
\label{zf2}
\eea
and we do not have any ghost fields because we directly work with the ghost determinant ${\rm det}(\frac{\delta G^d( Q_+)}{\delta \omega_+^e})$ in eq. (\ref{zfb}).

For the type I gauge transformation we have \cite{a,z}
\bea
&& T^d A'^{\lambda d}_+ = \Phi_+ T^dA^{\lambda d}_+ \Phi^{-1} +\frac{1}{ig} (\partial^\lambda \Phi_+)\Phi^{-1}_+, \nonumber \\
&& T^d Q'^{\lambda d}_+ = \Phi_+ T^dQ^{\lambda d}_+\Phi^{-1}_+
\label{zgt}
\eea
where the light-like gauge link or the light-like eikonal line in the fundamental representation of SU(3) is given by \cite{nnr,npp,npp1}
\bea
\Phi_+(x) =e^{igT^d \omega^d_+(x)}={\cal P}e^{-igT^d\int_0^\infty d\tau l \cdot A^d_+(x+\tau l)},~~~~~~~~~~l^2=0
\label{cmglf}
\eea
where $l^\lambda$ is the light-like four-velocity.

In this paper we will use the generating functionals from eqs. (\ref{zfnq}) and (\ref{zfb}) in the path integral formulation to prove the factorization of infrared divergences in the $\chi_{cJ}$ production from the color singlet $c{\bar c}$ pair in non-equilibrium QCD at RHIC and LHC at all orders of coupling constant.

\section{ Infrared divergences in $\chi_{cJ}$ production from color singlet $C{\bar C}$ pair}

The non-canceling infrared divergences were found in the higher order pQCD calculation of the annihilation of heavy quark-antiquark pair to light partons in the hadronic decay of the color singlet P-wave heavy quarkonium \cite{b}. For example, in the partonic processes \cite{b}
\bea
\chi_{cJ} \rightarrow q{\bar q}g,~~~~~~~~~~~~~~~~~h_c \rightarrow ggg
\label{cd}
\eea
of the hadronic decay of $\chi_{cJ}$ and $h_c$ respectively, one finds the non-canceling infrared divergences due to real soft gluons (eikonal gluons) emission/absorption \cite{b,br,cc}.

Now let us discuss the hadroproduction of $\chi_{cJ}$ from color singlet $c{\bar c}$ pair at high energy colliders. If the factorization theorem is valid \cite{sc,sc1,sg,nnr,npp,npp1} then the $\chi_{cJ}$ production from the color singlet $c{\bar c}$ pair at high energy colliders is given by
\bea
d\sigma_{pp \rightarrow \chi_{cJ} +X(P_T)} = \sum_{k,j}\int dx_1 dx_2 f_{k/p}(x_1,Q) f_{j/p}(x_2,Q) ~d{\hat \sigma}_{kj \rightarrow C{\bar C}[^3P_J] +X(P_T)} ~<0|{\cal O}_{\chi_{cJ}}|0>
\label{mcssip}
\eea
where $d{\hat \sigma}_{kj \rightarrow C{\bar C}[^3P_J] +X(P_T)}$ is the partonic level cross section for the $c{\bar c}$ production in $^3P_J$ state. This partonic level cross section can be calculated by using pQCD where $k,j=q,{\bar q},g$. The parton distribution function $f_{k/p}(x,Q)$ of the parton $k$ inside the proton $p$ is a non-perturbative quantity in QCD. The non-perturbative matrix element of $\chi_{cJ}$ production from the color singlet $c{\bar c}$ pair is denoted by $<0|{\cal O}_{\chi_{cJ}}|0>$.

As mentioned above the non-canceling infrared divergences were found in the hadronic decay of the color singlet P-wave heavy quarkonium \cite{b,br,cc}. Similarly, the non-canceling infrared divergences were also found in the hadroproduction of the color singlet P-wave heavy quarkonium \cite{cc}.

Note that for S-wave and P-wave color singlet heavy quarkonium the infrared divergences occur due to coulomb gluon and eikonal gluon exchanges. The infrared divergence due to Coulomb gluon exchange is analogous to the infrared divergence due to the Coulomb photon exchange in QED, see \cite{ha}. This Coulomb gluon infrared divergence is also known as the $\frac{1}{v} \rightarrow \infty$ divergence where $v$ is the relative velocity of the heavy quark-antiquark
which is normally absorbed into the normalization of the bound state wave function \cite{b} similar to that in QED \cite{ha}.

In case of $j/\psi$ production the infrared divergences due to the eikonal gluons
interacting with charm quark exactly cancel with the corresponding infrared divergences associated with the charm antiquark \cite{b}. Hence there is no uncanceled infrared divergences due to eikonal gluons exchange in case of $j/\psi$ production. That is why there are no gauge links in the definition of the $j/\psi$ wave function \cite{np}.

However, in case of $\chi_{cJ}$ production the non-canceling infrared divergences occur due to the eikonal gluons \cite{b}. At NLO in coupling constant the non-canceling infrared divergence due to the eikonal gluons exchange is found in the quark-antiquark fusion process \cite{cc}
\bea
q {\bar q} \rightarrow \chi_{cJ} g.
\label{qqb}
\eea
Because of the existence of these non-canceling infrared divergences, we have shown in \cite{np} that the gauge links are necessary in the definition of the color singlet
P-wave non-perturbative matrix element of the heavy quarkonium production. These gauge links make the non-perturbative matrix element gauge invariant and cancel these non-canceling infrared divergences.

Hence the correct definition of the non-perturbative matrix element of the $\chi_{c0}$ production from color singlet $c{\bar c}$ pair at high energy colliders which is consistent with the factorization of infrared divergences at all orders in coupling constant in QCD is given by \cite{np}
\bea
<0|{\cal O}_{\chi_{c0}}|0>= <0|\zeta^\dagger \Phi {\bar {\bf \nabla}} \Phi^\dagger \xi a^\dagger_{\chi_{c0}} \cdot a_{\chi_{c0}} \xi^\dagger \Phi {\bar {\bf \nabla}} \Phi^\dagger \zeta |0>
\label{mpwc}
\eea
where $\zeta$ ($\xi$) is the two component Dirac spinor field that creates (annihilates) a heavy quark and
\bea
\zeta^\dagger \Phi {\bar {\bf \nabla}} \Phi^\dagger \xi  = \zeta^\dagger \Phi (\vec{\nabla} \Phi^\dagger \xi )-(\vec{\nabla }\Phi^\dagger \zeta)^\dagger \Phi^\dagger \xi.
\label{mlrag}
\eea
In eq. (\ref{mpwc}) the $a^\dagger_{\chi_{c0}}$ is the creation operator of the $\chi_{c0}$, the $<0|{\cal O}_{\chi_{c0}}|0>$ is evaluated at the origin and
\bea
\Phi(x) ={\cal P}e^{-igT^d\int_0^\infty d\tau l \cdot A^d(x+\tau l)},~~~~~~~~~~l^2=0
\label{mglf}
\eea
is the light-like gauge link or the light-like eikonal line in the fundamental representation of SU(3).

\section{ Infrared divergence due to eikonal gluon and the SU(3) pure gauge background field }

As mentioned earlier the real gluon emission/absorption is the source of the non-canceling infrared divergences in case of P-wave heavy quarkonium production/deacy \cite{b,br,cc}. In this section we will briefly discuss the infrared divergence due to real gluon emission/absorption which can be described by eikonal Feynman rules in QCD. Let us first discuss the eikonal Feynman rules in QED before proceeding to QCD as the eikonal Feynman rules in QCD is similar to that in QED.

In QED the Feynman diagram contribution for an electron emitting a real photon is given by \cite{gm}
\bea
&&\frac{1}{{\not r} -{\not k} -m} {\not \epsilon}(k) u(r) = -\frac{r \cdot \epsilon(k)}{r \cdot k} u(r) +\frac{{\not k} {\not \epsilon}(k)}{2 r \cdot k}u(r)
\label{gra}
\eea
where $r^\lambda$ ($k^\lambda$) is the momentum of electron (photon). Eq. (\ref{gra}) has both eikonal part
\bea
\frac{r \cdot \epsilon(k)}{r \cdot k}u(r) \rightarrow \infty ~~~~~~~~~~~{\rm when}~~~~~~~~~k^\lambda \rightarrow 0
\label{gr1}
\eea
and the non-eikonal part
\bea
\frac{{\not k} {\not \epsilon}(k)}{2 r \cdot k}u(r) \rightarrow {\rm finite} ~~~~~~~~~~~{\rm when}~~~~~~~~~k^\lambda \rightarrow 0.
\label{gr2}
\eea
The eikonal part is the source of the infrared divergence as eq. (\ref{gr1}) diverges in the infrared limit $k^\lambda \rightarrow 0$. The non-eikonal part in eq. (\ref{gr2}) does not diverge in the infrared limit $k^\lambda \rightarrow 0$. This implies that the infrared divergence due to the emission of real photon from the electron can be studied by using only the eikonal term $\frac{r \cdot \epsilon(k)}{r \cdot k} u(r)$ without taking into account the non-eikonal term $\frac{{\not k} {\not \epsilon}(k)}{2 r \cdot k}u(r)$ in the Feynman diagram contribution in eq. (\ref{gra}).

Now we will show that the study of the infrared divergences due to the eikonal photons
at all order in coupling constant in QED can be enormously simplified when the electron is light-like ($r^2=0$).
The effective lagrangian density of the photon in the presence of current density $K^\lambda(x)$ in quantum field theory is given by \cite{nnr}
\bea
\int d^4x {\cal L}_{eff}(x) = -i~{\rm ln} <0|0>_K=-i ~{\rm ln}[\frac{Z[K]}{Z[0]}] = -\frac{1}{2}\int d^4x K^\lambda(x) \frac{1}{\partial^2} K_\lambda(x)
\label{gr3}
\eea
where the generating functional $Z[K]$ in the path integral formulation involving the photon field $Q^\lambda(x)$ is given by
\bea
Z[K] = \int [dQ] e^{i \int d^4x [-\frac{1}{4} [\partial_\delta Q_\lambda(x) - \partial_\lambda Q_\delta(x)][\partial^\delta Q^\lambda(x) - \partial^\lambda Q^\delta(x)] -\frac{1}{2 \alpha} (\partial_\lambda Q^{\lambda })^2+K_\lambda(x) Q^\lambda(x)]}.
\label{gr4}
\eea

From eq. (\ref{gr1}) the eikonal contribution
\bea
e \int \frac{d^4k}{(2\pi)^4} \frac{l_\lambda Q^\lambda(k)}{l \cdot k + i\epsilon} = -i\int d^4x Q^\lambda(x) K_\lambda(x)
\eea
gives the eikonal current density
\bea
K^\lambda(x) = e \int_0^\infty d\tau l^\lambda \delta^{(4)}(x-l\tau)
\label{gr5}
\eea
where $l^\lambda$ is the light-like four-velocity ($l^2=0$) of the electron.

Using eq. (\ref{gr5}) in (\ref{gr3}) we find that
\bea
{\cal L}_{eff}(x) = \frac{[el^2]^2}{[\sqrt{2}(l \cdot x)^2]^2}=0,~~~~~{\rm when}~~~~~~l\cdot x \neq 0,~~~~~~l^2=0.
\label{gr6}
\eea
From eq. (\ref{gr6}) we find that the light-like eikonal current produces pure gauge field in quantum field theory at all space-time points except at the positions perpendicular to the direction of motion of the charge at the time of closest approach, a result which agrees with the classical mechanics \cite{sc1,g1,g2}.

Hence we find from eq. (\ref{gr6}) that the calculation of infrared divergences due to the real photons emission from the light-like electron can be simplified by using the pure gauge field in QED. This can also be seen from Grammer-Yennie approximation \cite{gm} as follows. We write the photon polarization as the sum of the transverse (physical) polarization plus the longitudinal (pure gauge) polarization to find \cite{gm}
\bea
\epsilon^\lambda(k) = \epsilon^\lambda_{physical}(k)+\epsilon^\lambda_{pure~gauge}(k)
\label{gr9}
\eea
where
\bea
\epsilon^\lambda_{physical}(k)=[\epsilon^\lambda(k) - k^\lambda \frac{r \cdot \epsilon(k)}{r \cdot k}]
\label{g10}
\eea
which contributes to the physical (finite) cross section and
\bea
\epsilon^\lambda_{pure~gauge}(k)= k^\lambda \frac{r \cdot \epsilon(k)}{r \cdot k}
\label{gr11}
\eea
which does not contribute to the physical (finite) cross section but contributes to the the infrared divergence. This can be explicitly seen by using eq. (\ref{gr9}) in the eikonal part in eq. (\ref{gra}) to find
\bea
\frac{r \cdot \epsilon(k)}{r \cdot k}u(r)=\frac{r \cdot \epsilon_{pure~gauge}(k)}{r \cdot k} u(r) \rightarrow \infty ~~~~~~~~~~~{\rm when}~~~~~~~~~k^\lambda \rightarrow 0,
\label{gr1a}
\eea
\bea
\frac{r \cdot \epsilon_{physical}(k)}{r \cdot k}u(r) =0,
\label{gr1b}
\eea
and in the non-eikonal part in eq. (\ref{gra}) to find
\bea
\frac{{\not k} {\not \epsilon}(k)}{2 r \cdot k} u(r)=\frac{{\not k} {\not \epsilon}_{physical}(k)}{2 r \cdot k}u(r) \rightarrow {\rm finite} ~~~~~~~~~~~{\rm when}~~~~~~~~~k^\lambda \rightarrow 0
\label{gr2a}
\eea
and
\bea
\frac{{\not k} {\not \epsilon}_{pure~gauge}(k)}{2 r \cdot k}u(r) =0.
\label{gr2b}
\eea

From eq. (\ref{gr2}) the non-eikonal contribution
\bea
e\int \frac{d^4k}{(2\pi)^4} \frac{{\not k} {\not Q}(k)}{2 r \cdot k+ i\epsilon} = \int d^4x K(x) \cdot Q(x)
\eea
gives the non-eikonal current density
\bea
K^\lambda(x) = \frac{e}{2} \gamma^\delta \gamma^\lambda \int dw \frac{\partial}{\partial x^\delta} \delta^{(4)}(x-rw)
\label{gr7}
\eea
where $r^\lambda$ is light-like ($r^2=0$) or non-light-like ($r^2\neq 0$) momentum of the electron. Using eqs. (\ref{gr5}) and (\ref{gr7}) in eq. (\ref{gr3}) we find that the interaction between the (light-like or non-light-like) non-eikonal line with four-momentum $r^\lambda$ and the gauge field generated by the light-like eikonal line with four-velocity $l^\lambda$ ($l^2=0$) gives the interaction (effective) lagrangian density
\bea
{\cal L}_{eff}^{interaction}(x) =  \frac{l^2e^2[(r\cdot l)(r \cdot x) -r^2 l\cdot x ]}{2[(r \cdot x)^2 -r^2 x^2]^{\frac{3}{2}}}
=0,~~~~~~~~~~~~~~{\rm when}~~~~~~~~~~~~~l\cdot x \neq 0,~~~~~~~~~~r\cdot x\neq 0. \nonumber \\
\label{gr8}
\eea

From eq. (\ref{gr8}) we find that, in quantum field theory, the interaction between the non-eikonal line and the gauge field generated by the light-like eikonal line does not contribute to the interaction (effective) lagrangian density. Since the light-like eikonal line produces pure gauge field in quantum field theory (see eq. (\ref{gr6})) we find from eqs. (\ref{gr8}) and (\ref{gr2b}) that the light-like eikonal line does not modify the finite physical cross section.

Hence we find from eqs. (\ref{gr6}), (\ref{gr8}), (\ref{gr1a}), (\ref{gr1b}), (\ref{gr2a}) and (\ref{gr2b}) that the study of infrared divergences in QED due to real photons emission from the light-like electron can be enormously simplified by using the pure gauge field without modifying the finite value of the cross section.

We have shown in eqs. (\ref{gr6}) and (\ref{gr8}) that the light-like electron produces pure gauge field in QED. This result in QED agrees with classical mechanics \cite{sc1,g1,g2}. Hence we find that the infrared divergences at all orders in coupling constant due to the real photons emission from the light-like electron in quantum field theory can be studied by using the path integral formulation of the background field method of quantum field theory in the presence of pure gauge background field \cite{tc,nd,nnr,npp,npp1}.

In QED the U(1) pure gauge field $A^\lambda(x)$ is given by $A^\lambda(x)=\partial^\lambda \omega(x)$ and in QCD the SU(3) pure gauge field $A^{\lambda d}(x)$ is given by \cite{nnr,npp,npp1}
\bea
T^dA^{\lambda d}(x) = \frac{1}{ig} [\partial^\lambda \Phi(x)]\Phi^{-1}(x)
\label{pgq}
\eea
where $\Phi(x)$ is the light-like gauge link or the light-like eikonal line in the fundamental representation of SU(3) given by eq. (\ref{mglf}).

\section{ Proof of factorization of $\chi_{cJ}$ production in non-equilibrium QCD at RHIC and LHC in Color Singlet Mechanism}

As discussed in section IV the infrared divergences due to the exchange of eikonal gluons with the light-like parton in QCD can be studied by using the path integral formulation of the background field method of QCD in the presence of SU(3) pure gauge background field as given by eq. (\ref{pgq}) \cite{nnr,npp,npp1}. Note that the path integral technique is suitable to study the properties of the non-perturbative quantities in QCD. It should be mentioned here that the properties of a non-perturbative function may not always be correctly studied by using the perturbative method no matter how many orders of perturbation theory is used. Take, for example, a non-perturbative function
\bea
f(g) =e^{-\frac{1}{g^2}}.
\label{npf}
\eea
The Taylor series at $g=0$ gives $f(g)=0$ to all all orders in perturbation theory but $f(g)\neq 0$ for $g\neq 0$.

Having considered the points mentioned above, one should note that perturbative QCD entered a new phase when the cancelation of the leading-order (LO) renormalons between the QCD potential and the pole masses of quark and antiquark was discovered (see for example \cite{1}). Convergence of the perturbative series improved dramatically and much more accurate perturbative predictions became available. Hence, in some later works (see, for example, \cite{2}) it was shown that perturbative predictions in QCD agree well with phenomenological QCD results (determined from heavy quarkonium spectroscopy) and lattice QCD calculations. For recent developments on color potential produced by the color charge of the quark, see \cite{g1,g2}.

In this paper we will use the path integral formulation of the background field method of QCD to predict the correct definition of the non-perturbative matrix element of the $\chi_{cJ}$ production from color singlet $c{\bar c}$ pair in non-equilibrium QCD which is gauge invariant and is consistent with the factorization of infrared divergences at all orders in coupling constant.

In the closed-time path integral formulation the generating functional in non-equilibrium QCD is given by eq. (\ref{zfnq}). Hence from eq. (\ref{zfnq}) we find
that the heavy quark-antiquark non-perturbative correlation function of the type $<in|{\bar \Psi}_r(x') \Psi_r(x') {\bar \Psi}_s(x'') \Psi_s(x'')|in>$ in non-equilibrium QCD is given by \cite{g,c,a,mt}
\bea
&&<in|{\bar \Psi}_r(x') {\bar{\bf \nabla}}_{x'} \Psi_r(x') \cdot {\bar \Psi}_s(x'') {\bar{\bf \nabla}}_{x''} \Psi_s(x'')|in>\nonumber \\
&&
=\int [dQ_+] [dQ_-]\Pi_{k=1}^3[d{\bar \psi}_{k+}] [d{\bar \psi}_{k-}] [d \psi_{k+} ] [d\psi_{k-}]~[d{\bar \Psi}_{+}] [d{\bar \Psi}_{-}] [d \Psi_{+} ] [d\Psi_{-}]\nonumber \\
&&
\times {\bar \Psi}_r(x') {\bar{\bf \nabla}}_{x'} \Psi_r(x') \cdot {\bar \Psi}_s(x'') {\bar{\bf \nabla}}_{x''} \Psi_s(x'')\times {\rm det}(\frac{\delta \partial_\lambda Q_+^{\lambda d}}{\delta \omega_+^e})\times {\rm det}(\frac{\delta \partial_\lambda Q_-^{\lambda d}}{\delta \omega_-^e}) \nonumber \\
&&{\rm exp}[i\int d^4x {\bf \{}-\frac{1}{4}{F^d}_{\lambda \delta}^2[Q_+]+\frac{1}{4}{F^d}_{\lambda \delta}^2[Q_-] -\frac{1}{2 \alpha}(\partial_\lambda Q_+^{\lambda d })^2+\frac{1}{2 \alpha} (\partial_\lambda Q_-^{\lambda d })^2\nonumber \\
&&+\sum_{k=1}^3{\bar \psi}_{k+}  [i\gamma^\lambda \partial_\lambda -m_k +gT^d\gamma^\lambda Q^d_{\lambda +}]  \psi_{k+} -\sum_{k=1}^3{\bar \psi}_{k-}  [i\gamma^\lambda \partial_\lambda -m_k +gT^d\gamma^\lambda Q^d_{\lambda -}]  \psi_{k-}\nonumber \\
&&+{\bar \Psi}_{+}  [i\gamma^\lambda \partial_\lambda -M +gT^d\gamma^\lambda Q^d_{\lambda +}]  \Psi_{+}-{\bar \Psi}_{-}  [i\gamma^\lambda \partial_\lambda -M +gT^d\gamma^\lambda Q^d_{\lambda -}]  \Psi_{-}{\bf \}}]\nonumber \\
&&  \times <Q_+,\psi_{1+},{\bar \psi}_{1+},\psi_{2+},{\bar \psi}_{2+},\psi_{3+},{\bar \psi}_{3+},\Psi_+,{\bar \Psi}_+,0|~\rho~|0,{\bar \psi}_{1-},\psi_{1-},{\bar \psi}_{2-},\psi_{2-},{\bar \psi}_{3-},\psi_{3-},\nonumber \\
&&{\bar \Psi}_-,\Psi_-,Q_->
\label{avg1i}
\eea
where $r,s=+,-$ are the closed-time path indices in non-equilibrium QCD (the repeated closed-time path indices $r,s$ in eq. (\ref{avg1i}) are not summed) and $|in>$ is the ground state in non-equilibrium QCD.

In the closed-time path integral formulation in non-equilibrium the generating functional in the background field method of QCD is given by eq. (\ref{zfb}). Hence from eq. (\ref{zfb}) we find that the heavy quark-antiquark nonequilibrium-nonperturbative correlation function of the type $<in|{\bar \Psi}_r(x') \Psi_r(x') {\bar \Psi}_s(x'') \Psi_s(x'')|in>_A$ in the background field method of QCD is given by \cite{g,c,t,a,z}
\bea
&&<in|{\bar \Psi}_r(x') {\bar{\bf \nabla}}_{x'} \Psi_r(x') \cdot {\bar \Psi}_s(x'') {\bar{\bf \nabla}}_{x''} \Psi_s(x'')|in>_A\nonumber \\
&&
=\int [dQ_+] [dQ_-]\Pi_{k=1}^3[d{\bar \psi}_{k+}] [d{\bar \psi}_{k-}] [d \psi_{k+} ] [d\psi_{k-}]~[d{\bar \Psi}_{+}] [d{\bar \Psi}_{-}] [d \Psi_{+} ] [d\Psi_{-}]\nonumber \\
&&
\times {\bar \Psi}_r(x') {\bar{\bf \nabla}}_{x'} \Psi_r(x') \cdot {\bar \Psi}_s(x'') {\bar{\bf \nabla}}_{x''} \Psi_s(x'')\times {\rm det}(\frac{\delta G^d( Q_+)}{\delta \omega_+^e})\times {\rm det}(\frac{\delta G^d(Q_-)}{\delta \omega_-^e}) \nonumber \\
&&{\rm exp}[i\int d^4x {\bf \{}-\frac{1}{4}{F^d}_{\lambda \delta}^2[Q_++A_+]+\frac{1}{4}{F^d}_{\lambda \delta}^2[Q_-+A_-] -\frac{1}{2 \alpha}(G^d(Q_+))^2+\frac{1}{2 \alpha} (G^d(Q_-))^2\nonumber \\
&&+\sum_{k=1}^3{\bar \psi}_{k+}  [i\gamma^\lambda \partial_\lambda -m_k +gT^d\gamma^\lambda (Q+A)^d_{\lambda +}]  \psi_{k+} -\sum_{k=1}^3{\bar \psi}_{k-}  [i\gamma^\lambda \partial_\lambda -m_k +gT^d\gamma^\lambda (Q+A)^d_{\lambda -}]  \psi_{k-}\nonumber \\
&&+{\bar \Psi}_{+}  [i\gamma^\lambda \partial_\lambda -M +gT^d\gamma^\lambda (Q+A)^d_{\lambda +}]  \Psi_{+}-{\bar \Psi}_{-}  [i\gamma^\lambda \partial_\lambda -M +gT^d\gamma^\lambda (Q+A)^d_{\lambda -}]  \Psi_{-}{\bf \}}]\nonumber \\
&& <Q_++A_+,\psi_{1+},{\bar \psi}_{1+},\psi_{2+},{\bar \psi}_{2+},\psi_{3+},{\bar \psi}_{3+},\Psi_+,{\bar \Psi}_+,0|~\rho~|0,{\bar \psi}_{1-},\psi_{1-},{\bar \psi}_{2-},\psi_{2-},{\bar \psi}_{3-}\nonumber \\
&&,\psi_{3-},{\bar \Psi}_-,\Psi_-,Q_-+A_->.
\label{avg1ii}
\eea
From eq. (\ref{avg1ii}) we find
\bea
&&<in|{\bar \Psi}(x') \Phi(x') {\bar{\bf \nabla}}_{x'} \Phi^\dagger(x') \Psi(x') \cdot {\bar \Psi}(x'') \Phi(x'') {\bar{\bf \nabla}}_{x''} \Phi^\dagger(x'') \Psi(x'')|in>_A\nonumber \\
&&
=\int [dQ_+] [dQ_-]\Pi_{k=1}^3[d{\bar \psi}_{k+}] [d{\bar \psi}_{k-}] [d \psi_{k+} ] [d\psi_{k-}]~[d{\bar \Psi}_{+}] [d{\bar \Psi}_{-}] [d \Psi_{+} ] [d\Psi_{-}]\nonumber \\
&&
\times {\bar \Psi}(x') \Phi(x') {\bar{\bf \nabla}}_{x'} \Phi^\dagger(x') \Psi(x') \cdot {\bar \Psi}(x'') \Phi(x'') {\bar{\bf \nabla}}_{x''} \Phi^\dagger(x'') \Psi(x'')\times {\rm det}(\frac{\delta G^d( Q_+)}{\delta \omega_+^e})\times {\rm det}(\frac{\delta G^d(Q_-)}{\delta \omega_-^e}) \nonumber \\
&&{\rm exp}[i\int d^4x {\bf \{}-\frac{1}{4}{F^d}_{\lambda \delta}^2[Q_++A_+]+\frac{1}{4}{F^d}_{\lambda \delta}^2[Q_-+A_-] -\frac{1}{2 \alpha}(G^d(Q_+))^2+\frac{1}{2 \alpha} (G^d(Q_-))^2\nonumber \\
&&+\sum_{k=1}^3{\bar \psi}_{k+}  [i\gamma^\lambda \partial_\lambda -m_k +gT^d\gamma^\lambda (Q+A)^d_{\lambda +}]  \psi_{k+} -\sum_{k=1}^3{\bar \psi}_{k-}  [i\gamma^\lambda \partial_\lambda -m_k +gT^d\gamma^\lambda (Q+A)^d_{\lambda -}]  \psi_{k-}\nonumber \\
&&+{\bar \Psi}_{+}  [i\gamma^\lambda \partial_\lambda -M +gT^d\gamma^\lambda (Q+A)^d_{\lambda +}]  \Psi_{+}-{\bar \Psi}_{-}  [i\gamma^\lambda \partial_\lambda -M +gT^d\gamma^\lambda (Q+A)^d_{\lambda -}]  \Psi_{-}{\bf \}}]\nonumber \\
&&  \times <Q_++A_+,\psi_{1+},{\bar \psi}_{1+},\psi_{2+},{\bar \psi}_{2+},\psi_{3+},{\bar \psi}_{3+},\Psi_+,{\bar \Psi}_+,0|~\rho~|0,{\bar \psi}_{1-},\psi_{1-},{\bar \psi}_{2-},\psi_{2-},{\bar \psi}_{3-},\psi_{3-}\nonumber \\
&&,{\bar \Psi}_-,\Psi_-,Q_-+A_->
\label{avg3i}
\eea
where $\Phi(x)$ is the light-like gauge link or the light-like eikonal line in the fundamental representation of SU(3) given by eq. (\ref{mglf}).

Since $Q$ is the integration variable inside the path integration we change the integration variable $Q \rightarrow Q-A$ in eq. (\ref{avg3i}) to find
\bea
&&<in|{\bar \Psi}_r(x') \Phi_r(x') {\bar{\bf \nabla}}_{x'} \Phi^\dagger_r(x') \Psi_r(x') \cdot {\bar \Psi}_s(x'') \Phi_s(x'') {\bar{\bf \nabla}}_{x''} \Phi^\dagger_s(x'') \Psi_s(x'')|in>_A\nonumber \\
&&
=\int [dQ_+] [dQ_-]\Pi_{k=1}^3[d{\bar \psi}_{k+}] [d{\bar \psi}_{k-}] [d \psi_{k+} ] [d\psi_{k-}]~[d{\bar \Psi}_{+}] [d{\bar \Psi}_{-}] [d \Psi_{+} ] [d\Psi_{-}]\nonumber \\
&&
\times {\bar \Psi}_r(x') \Phi_r(x') {\bar{\bf \nabla}}_{x'} \Phi^\dagger_r(x') \Psi_r(x') \cdot {\bar \Psi}_s(x'') \Phi_s(x'') {\bar{\bf \nabla}}_{x''} \Phi^\dagger_s(x'') \Psi_s(x'')\times {\rm det}(\frac{\delta G^d_f( Q_+)}{\delta \omega_+^e})\nonumber \\
&&\times {\rm det}(\frac{\delta G^d_f(Q_-)}{\delta \omega_-^e}) \times {\rm exp}[i\int d^4x {\bf \{}-\frac{1}{4}{F^d}_{\lambda \delta}^2[Q_+]+\frac{1}{4}{F^d}_{\lambda \delta}^2[Q_-] -\frac{1}{2 \alpha}(G^d_f(Q_+))^2+\frac{1}{2 \alpha} (G^d_f(Q_-))^2\nonumber \\
&&+\sum_{k=1}^3{\bar \psi}_{k+}  [i\gamma^\lambda \partial_\lambda -m_k +gT^d\gamma^\lambda Q^d_{\lambda +}]  \psi_{k+} -\sum_{k=1}^3{\bar \psi}_{k-}  [i\gamma^\lambda \partial_\lambda -m_k +gT^d\gamma^\lambda Q^d_{\lambda -}]  \psi_{k-}\nonumber \\
&&+{\bar \Psi}_{+}  [i\gamma^\lambda \partial_\lambda -M +gT^d\gamma^\lambda Q^d_{\lambda +}]  \Psi_{+}-{\bar \Psi}_{-}  [i\gamma^\lambda \partial_\lambda -M +gT^d\gamma^\lambda Q^d_{\lambda -}]  \Psi_{-}{\bf \}}]\nonumber \\
&&  \times <Q_+,\psi_{1+},{\bar \psi}_{1+},\psi_{2+},{\bar \psi}_{2+},\psi_{3+},{\bar \psi}_{3+},\Psi_+,{\bar \Psi}_+,0|~\rho~|0,{\bar \psi}_{1-},\psi_{1-},{\bar \psi}_{2-},\psi_{2-},{\bar \psi}_{3-},\psi_{3-},\nonumber \\
&&{\bar \Psi}_-,\Psi_-,Q_->
\label{avg3ia}
\eea
where from eqs. (\ref{zgf}) and (\ref{zgt}) we have
\bea
&& G^d_f(Q_+) =\partial_\lambda Q^{\lambda d}_+ + gf^{dba} A_{\lambda +}^b Q^{\lambda a}_+-\partial_\lambda  A^{\lambda d}_+, \nonumber \\
&& T^d Q'^{\lambda d}_+ = \Phi_+ T^dQ^{\lambda d}_+ \Phi^{-1}_+ +\frac{1}{ig} (\partial^\lambda \Phi_+)\Phi^{-1}_+.
\label{gt2i}
\eea
Since $Q$, $\psi$, ${\bar \psi}$, $\Psi$ and ${\bar \Psi}$ are integration variables inside the path integration we can change the unprimed integration variables to primed integration variables in eq. (\ref{avg3ia}) to find
\bea
&&<in|{\bar \Psi}_r(x') \Phi_r(x') {\bar{\bf \nabla}}_{x'} \Phi^\dagger_r(x') \Psi_r(x') \cdot {\bar \Psi}_s(x'') \Phi_s(x'') {\bar{\bf \nabla}}_{x''} \Phi^\dagger_s(x'') \Psi_s(x'')|in>_A\nonumber \\
&&
=\int [dQ'_+] [dQ'_-]\Pi_{k=1}^3[d{\bar \psi}'_{k+}] [d{\bar \psi}'_{k-}] [d \psi'_{k+} ] [d\psi'_{k-}]~[d{\bar \Psi}'_{+}] [d{\bar \Psi}'_{-}] [d \Psi'_{+} ] [d\Psi'_{-}]\nonumber \\
&&
\times {\bar \Psi}'_r(x') \Phi_r(x') {\bar{\bf \nabla}}_{x'} \Phi^\dagger_r(x') \Psi'_r(x') \cdot {\bar \Psi}'_s(x'') \Phi_s(x'') {\bar{\bf \nabla}}_{x''} \Phi^\dagger_s(x'') \Psi'_s(x'')\times {\rm det}(\frac{\delta G^d_f( Q'_+)}{\delta \omega_+^e}) \nonumber \\
&&\times {\rm det}(\frac{\delta G^d_f(Q'_-)}{\delta \omega_-^e}) \times {\rm exp}[i\int d^4x {\bf \{}-\frac{1}{4}{F^d}_{\lambda \delta}^2[Q'_+]+\frac{1}{4}{F^d}_{\lambda \delta}^2[Q'_-] -\frac{1}{2 \alpha}(G^d_f(Q'_+))^2+\frac{1}{2 \alpha} (G^d_f(Q'_-))^2\nonumber \\
&&+\sum_{k=1}^3{\bar \psi}'_{k+}  [i\gamma^\lambda \partial_\lambda -m_k +gT^d\gamma^\lambda Q'^d_{\lambda +}]  \psi'_{k+} -\sum_{k=1}^3{\bar \psi}'_{k-}  [i\gamma^\lambda \partial_\lambda -m_k +gT^d\gamma^\lambda Q'^d_{\lambda -}]  \psi'_{k-}\nonumber \\
&&+{\bar \Psi}'_{+}  [i\gamma^\lambda \partial_\lambda -M +gT^d\gamma^\lambda Q'^d_{\lambda +}]  \Psi'_{+}-{\bar \Psi}'_{-}  [i\gamma^\lambda \partial_\lambda -M +gT^d\gamma^\lambda Q'^d_{\lambda -}]  \Psi'_{-}{\bf \}}]\nonumber \\
&&  \times <Q'_+,\psi'_{1+},{\bar \psi}'_{1+},\psi'_{2+},{\bar \psi}'_{2+},\psi'_{3+},{\bar \psi}'_{3+},\Psi'_+,{\bar \Psi}'_+,0|~\rho~|0,{\bar \psi}'_{1-},\psi'_{1-},{\bar \psi}'_{2-},\psi'_{2-},{\bar \psi}'_{3-},\psi'_{3-},\nonumber \\
&&{\bar \Psi}'_-,\Psi'_-,Q'_->.
\label{avg4i}
\eea
The SU(3) pure gauge background field $A^{\lambda d}(x)$ given by eq. (\ref{pgq}). Using the background field $A^{\lambda d}(x)$ as the SU(3) pure gauge background field given by eq. (\ref{pgq}) we find from
\bea
\psi'_+(x) = \Phi_+(x) \psi_+(x)
\label{gt3i}
\eea
and from eq. (\ref{gt2i}) that \cite{nnr,npp,npp1}
\bea
&& [d{\bar \psi}'_{k +}] [d \psi'_{k+} ]=[d{\bar \psi}_{k+}] [d \psi_{k+} ],~~~~~~~~~~~~[dQ'_+] =[dQ_+],~~~~~~~~~~~[d{\bar \Psi}'_+] [d \Psi'_+ ]=[d{\bar \Psi}_+] [d \Psi_+ ],\nonumber \\
&& (G_f^d(Q'_+))^2 = (\partial_\lambda Q^{\lambda d}_+(x))^2,~~~~~~~~~~~{\rm det} [\frac{\delta G_f^d(Q'_+)}{\delta \omega^e_+}] ={\rm det}[\frac{ \delta (\partial_\lambda Q^{\lambda d}_+(x))}{\delta \omega^e_+}] \nonumber \\
&&{\bar \psi}'_{k+} [i\gamma^\lambda \partial_\lambda -m_k +gT^d\gamma^\lambda Q'^d_{\lambda + }] \psi'_{k+}={\bar \psi}_{k+} [i\gamma^\lambda \partial_\lambda -m_k +gT^d\gamma^\lambda Q^d_{\lambda +}] \psi_{k +}, \nonumber \\
&&{\bar \Psi}'_+ [i\gamma^\lambda \partial_\lambda -M +gT^d\gamma^\lambda Q'^d_{\lambda +}] \Psi'_\pm={\bar \Psi}_+ [i\gamma^\lambda \partial_\lambda -M +gT^d\gamma^\lambda Q^d_{\lambda +}]\Psi_+.
\label{gt4i}
\eea
At the initial time we are working in the frozen ghost formalism for the non-equilibrium QCD at the initial time \cite{g,c}. This implies from eqs. (\ref{gt2i}) and (\ref{gt3i}) that at the initial time the $<Q_+,\psi_{1+},{\bar \psi}_{1+},\psi_{2+},{\bar \psi}_{2+},\psi_{3+},{\bar \psi}_{3+},\Psi_+,{\bar \Psi}_+,0|~\rho~|0,{\bar \psi}_{1-},\psi_{1-},{\bar \psi}_{2-},\psi_{2-},{\bar \psi}_{3-},\psi_{3-},{\bar \Psi}_-,\Psi_-,Q_->$ in non-equilibrium QCD at the initial time is gauge invariant by definition, {\it i. e.}, \cite{npp1}
\bea
&&<Q'_+,\psi'_{1+},{\bar \psi}'_{1+},\psi'_{2+},{\bar \psi}'_{2+},\psi'_{3+},{\bar \psi}'_{3+},\Psi'_+,{\bar \Psi}'_+,0|~\rho~|0,{\bar \psi}'_{1-},\psi'_{1-},{\bar \psi}'_{2-},\psi'_{2-},{\bar \psi}'_{3-},\psi'_{3-},\nonumber \\
&&{\bar \Psi}'_-,\Psi'_-,Q'_->\nonumber \\
&&=<Q_+,\psi_{1+},{\bar \psi}_{1+},\psi_{2+},{\bar \psi}_{2+},\psi_{3+},{\bar \psi}_{3+},\Psi_+,{\bar \Psi}_+,0|~\rho~|0,{\bar \psi}_{1-},\psi_{1-},{\bar \psi}_{2-},\psi_{2-},{\bar \psi}_{3-},\psi_{3-},\nonumber \\
&&{\bar \Psi}_-,\Psi_-,Q_->.
\label{ddn}
\eea

From eqs. (\ref{gt4i}), (\ref{gt3i}), (\ref{ddn}), (\ref{avg4i}) and (\ref{avg1i}) we finally obtain
\bea
&&<in|{\bar \Psi}_r(x') {\bar{\bf \nabla}}_{x'} \Psi_r(x') a^\dagger_H \cdot a_H {\bar \Psi}_s(x) {\bar{\bf \nabla}}_x \Psi_s(x)|in>\nonumber \\
&&=   <in|{\bar \Psi}_r(x') \Phi_r(x'){\bar{\bf \nabla}}_{x'} \Phi^\dagger_r(x') \Psi_r(x') a^\dagger_H \cdot a_H {\bar \Psi}_s(x) \Phi_s(x) {\bar{\bf \nabla}}_x \Phi^\dagger_s(x) \Psi_s(x)|in>_A
\label{facti}
\eea
which proves the factorization of infrared divergences in $\chi_{cJ}$ production from color singlet $c{\bar c}$ pair in non-equilibrium QCD at all order in coupling constant where the light-like gauge link or the light-like eikonal line $\Phi_+(x)$ in the fundamental representation of SU(3) is given by
\bea
\Phi_+(x) ={\cal P}e^{-igT^d\int_0^\infty d\tau l \cdot A^d_+(x+\tau l)}.
\label{glii}
\eea

\section{Correct Definition of $\chi_{cJ}$ production in non-equilibrium QCD at RHIC and LHC in color singlet mechanism }

From eq. (\ref{facti}) we find that the correct definition of the gauge invariant non-perturbative matrix element of the $\chi_{c0}$ production from the color singlet $c{\bar c}$ pair in non-equilibrium QCD which is consistent with factorization of infrared divergences at all orders in coupling constant is given by
\bea
<in|{\cal O}_{\chi_{c0}}|in> = <in|\zeta^\dagger \Phi {\bar {\bf \nabla}} \Phi^\dagger \xi a^\dagger_{\chi_{c0}} \cdot a_{\chi_{c0}} \xi^\dagger \Phi {\bar {\bf \nabla}} \Phi^\dagger \zeta |in>.
\label{pwcfi}
\eea
Since the left hand side of eq. (\ref{facti}) is independent of the light-like four-velocity $l^\lambda$ we find that the long-distance behavior of the $\chi_{c0}$ non-perturbative matrix element $<in|{\cal O}_{\chi_{c0}}|in> =<in|\zeta^\dagger \Phi {\bar {\bf \nabla}} \Phi^\dagger \xi a^\dagger_{\chi_{c0}} \cdot a_{\chi_{c0}} \xi^\dagger \Phi {\bar {\bf \nabla}} \Phi^\dagger \zeta |in>$ in eq. (\ref{pwcfi}) in non-equilibrium QCD is independent of the light-like vector $l^\lambda$ used to define the light-like gauge link or the light-like eikonal line in eq. (\ref{glii}) at all orders in coupling constant.

\section{Conclusions}
Recently we have proved the factorization of NRQCD S-wave heavy quarkonium production at all orders in coupling constant. In this paper we have extended this to prove the factorization of infrared divergences in  $\chi_{cJ}$ production from color singlet $c{\bar c}$ pair in non-equilibrium QCD at RHIC and LHC at all orders in coupling constant. This can be relevant to study the quark-gluon plasma at RHIC and LHC.

\end{document}